\documentclass[12pt,sort&compress]{article}

\usepackage[dvips]{color}
\usepackage{amsmath}
\usepackage{graphicx} 
\usepackage[dvips]{color}
\usepackage{graphicx}
\usepackage{tabularx}
\usepackage{subfig}
\usepackage{cite}
\usepackage[section]{placeins}
\usepackage{hyperref}

\newcommand{\proba}{\mathrm{Prob}}

\newcommand{\signum}{\mathrm{sign}}

\newcommand{\Ai}{\mathrm{Ai}}

\newcommand{\cbar}{\bar{c}}
\newcommand{\aStep}{\mathtt{a}}

\begin{document} 
\title{Crossover  in the log-gamma  polymer from the replica  coordinate Bethe Ansatz}
\author{Pascal Grange\\
{\emph{Department of Mathematical Sciences}}\\
 {\emph{Xi'an Jiaotong-Liverpool University}}\\
{\emph{111 Ren'ai Rd, 215123 Suzhou, China}}\\
\normalsize{{\ttfamily{pascal.grange@xjtlu.edu.cn}}}}

\date{}
\maketitle
\vspace{1cm}
\begin{abstract}
The  coordinate Bethe Ansatz solution of the log-gamma polymer is extended to 
 boundary conditions with one fixed end and the other attached to one half of a one-dimensional lattice. 
The large-time limit is studied using a saddle-point approximation,
 and the cumulative distribution function of the  rescaled free energy of a long polymer is expressed as a Fredholm determinant.
 Scaling limits of the kernel are identified, leading to a crossover from the  GUE to the GOE Tracy--Widom 
 distributions. The continuum limit reproduces the crossover from droplet to flat initial 
 conditions of the Kardar--Parisi--Zhang equation.
\end{abstract}

\newpage

\tableofcontents

\section{Introduction and conclusions}
The Kardar--Parisi--Zhang (KPZ) equation \cite{KPZ,KPZRep}, a continuum model 
 of one-dimensional growth of an interface  in the presence  of noise, 
 can be solved exactly by mapping  the height 
 field of the interface to the free energy of a continuum directed-polymer model. The time evolution of the 
 integer moments of the partition function of this model, is given by   the Hamiltonian of the 
 one-dimensional Lieb--Liniger model of interacting bosons \cite{LL,BrunetDerrida1,BrunetDerrida2}, which is solvable by Bethe Ansatz methods 
 \cite{GaudinTraduction}.  These moments allow to reconstruct the probability density
 of the rescaled free energy for the most studied classes of boundary conditions  
 for the KPZ equation. In particular, the free energy is related  at large times to  Tracy--Widom \cite{TW,GOEKernel}
 distributions of the largest eigenvalue of large Gaussian random matrices, with classes
 depending on the boundary conditions, such as the Gaussian unitary ensemble (or GUE)
  for droplet boundary conditions and the Gaussian orthogonal ensemble (or GOE) for flat boundary conditions.\\

 An alternative approach to the solution of the continuum model (with 
 KPZ universality properties related to classes of boundary conditions  \cite{Rosso,Dotsenko,DotsenkoKlumow,CalabreseLeDoussalFlat,CalabreseLeDoussalLong,crossoverLD,IS1,IS2,
GueudreLeDoussal,131,132,133,Spohn1,Spohn2,Spohn3,Spohn4,Spohn5,SasamotoSpohn1,Quastel,Corwin11,LeDoussal,CalabreseLeDoussalQuench,sineGordon}),
   is the study of discrete models of directed polymers  (see \cite{theseThiery} for a recent review,
 and \cite{Povolotsky,TLDSquare,ThieryStat,TLDBeta,TLDTimeDep,Yor,HigherSpin,strictWeak,betaRandom} for more families of models and their classification,
 and \cite{ISDuality,BCS14} for rigorous solutions involving the replica approach).
  In particular, the homogeneous  log-gamma polymer, introduced by Sepp\"al\"ainen \cite{logGammaDef},
 defined by the distribution of Boltzmann weights on a lattice  is a model of up-right directed paths on a square lattice with multiplicative random weights
 distributed according to the homogeneous log-gamma model are considered. 
 The Boltzmann  weights (at finite temperature set to unity)
  are  independent identically-distributed variables with the the distribution: 
\begin{equation}\label{Pgamma}
 P_\gamma(\omega) d\omega = \frac{1}{\Gamma( \gamma )} \omega^{-1-\gamma} e^{-1/\omega} d\omega, 
\end{equation}
 with fixed parameter $\gamma > 0$. The mapping from the integer coordinates $(i,j)$ of the vertices 
 to space and time coordinates $(x,t)$ is defined by the relations $t=i+j-2$ and $x=(i-j)/2$. The up-right 
 constraint on the directed  polymer amounts to alllowing only jumps from $(x,t)$ to $\left(x \pm\frac{1}{2}, t+1\right)$.
  The  allowed values of the coordinate $x$ at even (resp. odd) times are therefore integer (resp. half-odd) numbers.\\

The partition function of the log-gamma polymer with both ends fixed was 
studied in \cite{logGammaTLD} by coordinate Bethe Ansatz techniques and related to the continuum directed polymer known to be mapped to 
the KPZ equation with droplet boundary conditions, using the replica solution of the Lieb--Liniger model (LL).  It  was also 
 solved by combinatorial methods \cite{Corwin}, and the generating function was 
 related to a Fredholm determinant \cite{Borodin} (see also \cite{BisiZygouras} for a rigorous approach to the model at 
 finite time). Using a symmetry argument the rescaled free energy of the polymer with one free
 end was related to the GOE Tracy--Widom distribution \cite{logGammaFlat}.
 In this paper we will  extend the Bethe Ansatz approach to the limit of a long polymer 
 with one end fixed at $(x,t)$ and the other end attached to one half of an infinite 
 one-dimensional lattice.\\

 We need to define a discrete analogue of the partition function of the continuum polymer model, 
studied (see \cite{CalabreseLeDoussalLong,crossoverLD}) with boundary conditions defined by means of a slope $w$ 
 and denoted by 
\begin{equation}\label{halfFlatLL}
Z^{LL}_w(x,t) := \int_{-\infty}^0 dy e^{wy} Z^{LL}(x,t|y,0) = \lim_{w'\rightarrow \infty}\left( \int_{-\infty}^0 e^{w y\theta( -y) - w' y\theta(y)} Z^{LL}(x,t|y,0)  \right),
\end{equation}
 where $Z^{LL}(x,t|y,0)$ is the partition function of the continuum polymer with both ends fixed, one at $(x,t)$ and one at $(y,0)$
 with a weight factor associated with each sector.
 As the free energy of the continuum directed 
 polymer is identified to the height function in the KPZ growth model,
 this special regime of the  shift is  referred to as half-flat.\\

 In the discrete setting, let us define the partition function $Z_w(x,t)$ of the directed polymer starting with a fixed end 
   as the sum  of the Boltzmann weights of the directed  paths $\phi$ with $t$ time steps, with exponential weights
 given to the space coordinate of the starting point:
\begin{equation}\label{halfFlat}
Z_w(x,t) := \sum_{y \in {\mathbf{Z}}_-}\sum_{\phi: (y,0) \rightarrow (x,t) } \prod_{(x',t')\in \phi} e^{wy} \omega_{x',t'},
\end{equation}
 with space and time restricted to  discrete values  by the constraint of up-right directed paths as explained above.
 Due to the random nature of the Boltzmann weights, this partition function is a random variable. The purpose of this paper is to characterise
 its probability law in the large-time limit and to identify two limits, one involving large positive positions and one involving
 large negative positions, in which the law of a suitable rescaled free energy approaches the GOE and GUE distributions, 
 in order to establish the existence of a crossover regime in the log-gamma polymer.\\

 Our derivation is conjectural to the extend that we adopt the analytic-continuation 
 prescription to complex values of the number of replicas (which is needed
 as the fat tail of the distribution of Boltzmann weights gives rise to divergences
 for replicas of order larger than $\gamma$), and the conjectural form of the
 norm of Bethe states proposed in \cite{logGammaTLD}. Thios system of states is assumed to be complete.
 We will introduce a scaling  parameter $\lambda$ and by means of a saddle-point approximation in the 
 time-evolution of replicated partition functions, relate it 
 to time by a scaling law, and introduce a rescaled space coordinate $\tilde{x}$ and a rescaled slope $\tilde{w}$ as follows:
 \begin{equation}
\lambda^3 = -\psi''\left( \frac{\gamma}{2}\right)\frac{t}{8},\;\;\;\;\;\;\;\;\;\tilde{x}=2\psi'\left( \frac{\gamma}{2}\right) \frac{x}{\lambda^2},\;\;\;\;\;\;\;\;
\tilde{w}=\lambda w,
 \end{equation}
 where $\psi$ is the Euler digamma function.
 The large-time limit is therefore captured by the large-$\lambda$ limit.
  Let us denote by  $\chi_1$ (resp. $\chi_2$) a random variable whose cumulative distribution function 
 is given by the GOE function $F_1$ (resp. the GUE function $F_2$). In the large-$\lambda$ limit,
 we will identify the two  identities in law:
\begin{equation}\label{GUE}
\log( Z_w(x,t) )+\psi\left( \frac{\gamma}{2}\right)t \underset{{\tilde{w} + \tilde{x}/8\to-\infty, \tilde{w}>0}}{\simeq}\lambda \chi_1 + \lambda\left( 4\frac{\tilde{w}^2}{\cbar^2} +  \tilde{w}\frac{\tilde{x}}{\cbar} \right),
\end{equation}
\begin{equation}\label{GOE}
\log( Z_w(x,t) )+\psi\left( \frac{\gamma}{2}\right)t \underset{\tilde{w} + \tilde{x}/8\to \infty}\simeq \lambda \chi_2 -8\frac{\psi'\left(\frac{\gamma}{2}\right)^2}{\psi''\left(\frac{\gamma}{2}\right)}
\frac{x^2}{4t},
\end{equation}
 where $\cbar = 4/(\gamma-1)$.\\

The structure of the paper is as follows. In Section \ref{reviewSection} we introduce 
notations for  the model and the quantities of interest, then review the coordinate Bethe Ansatz solution 
and the assumptions leading to an expression the generating function, organised as a string expansion.
   In Section 3 we work out the one-string contribution, identify scaling laws 
  from a saddle-point approximation of the time-evolution factor in the large-time limit.
 This step provides an expression for the one-string contribution as the trace of a kernel.
 Arguments from the solution of the Lieb--Liniger problem  in the continuum with half-flat
 boundary condition are carried over in Section 4 to the saddle-point approximation of 
 the discrete model, leading to an organisation of the generating function as a Fredholm determinant.
  By introducing a lattice spacing we  show that 
 the continuum limits of Eqs \ref{GUE} and \ref{GOE} (which inducees a scaling of time and the limit of large $\gamma$ parameter)
  reproduce the GOE and GUE limits of the crossover in the KPZ equation \cite{crossoverLD}.

\section{Review of the model and quantities of interest}\label{reviewSection}

\subsection{Generating function and starting formula in terms of moments}
  We can access the cumulative distribution function of the 
 free energy  through the large-$\lambda$ limit of the 
 following generating function $g_\lambda$ defined by 
\begin{equation}\label{generatingDef}
  g_\lambda( s )= \overline{ \exp\left(  -e^{-\lambda s} Z_w(x,t)  \right)} = \overline{\exp\left(- e^{-\lambda( s+ f ) }\right)},
\end{equation}
 where the overline denotes the average over the distribution of Boltzmann weights (the disorder described by Eq. \ref{Pgamma}), and the rescaled free energy  $f$  is defined as
\begin{equation}
 f := -\lambda^{-1}\log\left(  Z_w(x,t)\right).
\end{equation}
 Indeed, provided the parameter $\lambda$ can go to infinity together with time,
 the large-time cumulative distribution function of $f$ 
 can be accessed through the large-$\lambda$ limit of the generating function:
\begin{equation}
 \lim_{\lambda\to\infty} g_\lambda( s)= \overline{\theta(f+s )} = \proba( f>-s).
\end{equation}
 On the other hand, an expansion of the exponential function in Eq.  \ref{generatingDef}
 at fixed $\lambda$ gives  rise to a starting expression of the generating function 
 as a formal series in the moments of the distribution of the random variable $Z_w(x,t)$:
\begin{equation}\label{formalExpansion}
 g_\lambda^{mom} ( s)  := 1 + \sum_{n\geq 1} (-1)^n\frac{e^{-\lambda ns}}{n!}\overline{ Z_w(x,t)^n}
\end{equation}
 However, the  moments of  $Z_w(x,t)$ do not exist beyond a certain order depending
 on the value of the model parameter $\gamma$, as was appreciated in \cite{logGammaTLD} 
 in the model with two fixed ends and
 in \cite{logGammaFlat} in the model with one fixed end, by considering the moment of 
 order $n$ and time $0$. In the present model, we observe the same divergence at large orders:
\begin{equation}
 \overline{ Z_w(x,0)^n} = \overline{\sum_{y\in {\mathbf{Z}}_-}\delta_{x,y} e^{wy} \left( \omega_{x,0}\right)^n } = \frac{1}{\Gamma( \gamma )} \int_0^\infty w^{-1 + n - \gamma} e^{-1/w} dw 
= e^{wx} \frac{\Gamma( \gamma-n)}{\Gamma( \gamma)},
\end{equation}
 but the expression can  manifestly be extended to complex values of $n$.\\
 
Let us assume that the generating function can be 
  expressed as complex integrals using the Mellin representation of the 
  exponential function, and use the prescription given in \cite{logGammaTLD}, leading from the 
 starting formal series $g_\lambda^{mom}$ to the complex integral   
\begin{equation}\label{prescription}
\begin{split}
g_\lambda( s )
&= - \int P(Z_w)  \left( 
\int_{C} \frac{dm}{2i\pi \sin(\pi m)} \frac{1}{\Gamma( 1 + m ) } e^{-\lambda s m} Z_w^m\right) dZ_w\\
&=  \int_{C} \frac{dm}{2i\pi \sin(\pi m)} \frac{1}{\Gamma( 1 + m ) } e^{-\lambda s m} \overline{Z_w^m},\\
\end{split}
\end{equation}
where $C = -a + i\mathbf{R}$ with $a>0$, and that the expression $\overline{Z_w^m}$ can be extended to complex values
 of the order $m$, at all times.
%The formal sum over integer values  can be reorganised as a sum over numbers of 
%strings \cite{logGammaTLD}, assuming completeness of the system of Bethe--Bunet states:
% \begin{equation}\label{stringExpansion}
%g^{mom}_\lambda( u ) =: 1 + \sum_{n_s\geq 1} \frac{1}{n_s!} Z^{mom}(n_s,u),
%\end{equation}
% which induces the definition of the formal  $n_s$-string term $Z^{mom}(n_s,u)$, containing
% replicas at all orders (the dependence on the point $x$ and on $w$ has been dropped from the notation), to which the analytic-continuation prescription can be applied.

\subsection{Quantum-mechanical calculation of the moments of the partition function}
 By analogy with the mapping from directed polymer in the continuum to the Lieb--Liniger
 model, the moment of order $n$ of the  polymer with two fixed ends (one being 
 at vertex $y$ and time $t=0$)  was mapped in \cite{logGammaTLD} to a wave function $\psi_t$
  by the following definition:
 \begin{equation}\label{evolZ}
\overline{\prod_{i=1}^n Z(x_i, t|y,0) } =: 2^{nt}\left(\frac{\bar{c}}{4}\right)^{n(t+1)}\psi_t(x_1,\dots,x_n),\;\;\;\mathrm{with}\;\;\;\cbar:=\frac{4}{\gamma-1}.
\end{equation}
  The multiplicativity  of the Boltzmann weights,  expressed between times $t$ and $t+1$ along a directed polymer
 path,
 \begin{equation}  
 Z(x, t+1|y,0)= w_{x,t+1}\left( Z\left( x-\frac{1}{2}, t|y,0\right)  +  Z\left( x+\frac{1}{2},t|y,0\right)\right),
 \label{zEvol}
\end{equation}
induces a linear time-evolution equation for the wave function:
\begin{equation}  
\psi_{t+1}=H_n\psi_t.
\end{equation}
The expression of the Hamiltonian (or transfer matrix) $H_n$ was worked out in Section 4 of \cite{logGammaTLD}.
 It is the analogue for the log-gamma polymer of the Lieb--Liniger Hamiltonian in the continuum directed polymer.  In the next section we will
  use the results of the diagonalisation problem of the Hamiltonian:
\begin{equation}
  H_n \Psi_\mu(x_1,\dots, x_n) = \theta_\mu \Psi_\mu (x_1,\dots, x_n).
\label{Schroedinger}
 \end{equation}
  Given a complete system of eigenfunctions labelled by $\mu$, with associated eigenvalues $\theta_\mu$, and the notation 
 $\Psi_\mu (x_1,\dots,x_n) = \langle x_1,\dots,x_n| \mu\rangle$,
 inserting projectors onto these eigenstates into the definition of $Z_w$ in Eq. \ref{halfFlat} yields the expression (up to a complex 
 conjugation that was also taken in \cite{crossoverLD} and has no effect on the l.h.s. which is a real quantity): 
\begin{equation} \label{momentOrdern}
\begin{split}
 \overline{ {Z_w(x,t)}^n}&= 2^{nt}\left(\frac{\bar{c}}{4}\right)^{n(t+1)}  \sum_{y_1,\dots, y_n \in \{0,\dots, L-1 \}}e^{w\sum_{p=1}^n y_p}
\langle y_1,\dots,y_n|H_n^t|x_1,\dots,x_n\rangle \\
 =& 2^{nt}\left(\frac{\bar{c}}{4}\right)^{n(t+1)}  \sum_\mu \sum_{y_1,\dots, y_n \in \{0,\dots, L-1 \}}e^{w\sum_{p=1}^n y_p}
\langle y_1,\dots,y_n| \mu\rangle\frac{\theta_\mu^t}{\langle \mu|\mu\rangle}\langle \mu | x_1,\dots,x_n\rangle \\
=&2^{nt}\left(\frac{\bar{c}}{4}\right)^{n(t+1)} \sum_\mu \Psi^\ast_\mu (x,\dots, x)\theta_\mu^t \frac{1}{\langle \mu | \mu \rangle}  \left( \sum_{y_1,\dots, y_n \in \{0,\dots, L-1 \}}e^{w\sum_{p=1}^n y_p}\Psi_\mu (y_1,\dots, y_n)\right).
\end{split}
\end{equation}

\subsection{Bethe Ansatz solution in the large-volume limit}
 Eigenfunctions of the operator $H_n$ can be expressed  by coordinate Bethe Ansatz techniques
 due to Brunet. They consist of a superposition of plane waves parametrised by rapidities denoted by $(\lambda_1,\dots,\lambda_n)$:
% In the limit of large $L$ (the thermodynamic limit), the 
%  {\emph{tangents}} of these rapidities can be grouped into strings are regularly spaced in the complex plane:
\begin{equation}
 \Psi_\mu(x_1,\dots,x_n)= \sum _{\sigma \in \mathcal{S}_n} A_\sigma \prod_{i=1}^n e^{i\lambda_{\sigma(\alpha)} x_\alpha},
 \label{BetheBrunet}
\end{equation}
but weighted  by symmetry factors written in terms of the tangents of the rapidities.
\begin{equation}
 A_\sigma = \prod_{1 \leq \alpha < \beta \leq n } \left(   1  +\frac{\bar{c}}{2}\frac{\signum( x_\beta - x_\alpha + 0^+)}{t_{\sigma(\alpha)} - t_{\sigma(\beta)}} \right), \;\;\;\;\;\;t_\alpha = i \tan\left( \frac{\lambda_\alpha}{2}\right),
\label{Asigma}
\end{equation}
 with the  notation
\begin{equation}
 \cbar = \frac{4}{\gamma-1}.
\label{oddMapping}
\end{equation}
  This Ansatz implies that the plane waves can be expressed in terms of the family of parameters $(t_\alpha)_{ 1 \leq \alpha \leq n}$:
\begin{equation}
 z_\alpha = e^{i\lambda_\alpha} = \frac{ 1 + t_\alpha}{ 1 - t_\alpha}.
\label{ttoz}
\end{equation}
 In the special case  $n=2$, working out the eigenvalues $\theta_\mu$ in the 
 time-evolution problem of  Eq. \ref{evolZ} yields $\theta_{\mu,(n=2)} = 
\frac{1}{4}\sum_{\alpha,\beta \in \left\{ -\frac{1}{2},\frac{1}{2}\right\}} z_1^\alpha z_2^\beta = \frac{1}{4}\prod_{\alpha=1}^2(z_\alpha^{\frac{1}{2}} + z_\alpha^{-\frac{1}{2}})$, which generalises at higher orders,  yielding an  
 expression of the eigenvalues in terms of the families of parameters $(t_\alpha)_{1\leq \alpha \leq n}$ only:
\begin{equation}\label{thetaMuExpr}
 \theta_\mu =\prod_{\alpha=1}^n \frac{(z_\alpha^{\frac{1}{2}} + z_\alpha^{-\frac{1}{2}})}{2}= \left(\prod_{\alpha=1}^n\frac{1}{1-t_\alpha^2} \right)^{\frac{1}{2}}.
 \end{equation}

Moreover, imposing $L$-periodic boundary conditions yields the following generalisation of the Bethe equations
\begin{equation}
e^{i\lambda_\alpha L}  = \prod_{\beta\neq \alpha} \frac{ 2t_\alpha - 2t_\beta +\bar{c} }{2t_\beta - 2t_\alpha -\bar{c} },\;\;\;\;\; \;\;\alpha\in \{1,\dots, n\}.
 \label{BetheEquations}
\end{equation}
 In the thermodynamic limit (of a large number $L$ of sites), when expressed in terms of the
 tangents of the rapidities (and not in terms of the rapidities themselves as in the 
 Lieb--Liniger model)  are arranged in {\emph{strings}} in the complex plane, up to corrections vanishing exponentially\footnote{All the results of this paper
 are derived in the thermodynamic limit.} at large $L$.  Indeed, if the rapidity $\lambda_\alpha$ 
 has a strictly positive imaginary part, the l.h.s. of Eq. \ref{BetheEquations} vanishes eponentially at $L$,
 which implies that one of the factors in the numerator of the r.h.s. must be zero in the thermodynamic limit. This implies that there exists 
 an index $\beta$ such that $2t_\beta = 2t_\alpha - \cbar$. Iterating this procedure (and repeating it 
 using the denominator of the r.h.s in the case of a negative imaginary part of the rapidity) yields
 a string of parameters, specified by $m_j$ regularly spaces values on a horizontal segment 
 in the complex plane:
\begin{equation}
 t_\alpha = t_{j,a} = i\frac{k_j}{2} + \frac{\bar{c}}{4}( m_j + 1 - 2a ),\;\;\;a\in \{1,\dots, m_j\}.
\label{talpha}
\end{equation}
 As such a  string is invariant by complex conjugation, the eigenfunctions 
 at identical values of the argument needed to express the moments of the partition function read,
 for a system of $n_s$ strings labelled by the integer $j$, each with $m_j$ rapidities and an imaginary part given by $ik_j/2$:
\begin{equation}\label{PsiMuGamma}
\begin{split}
 \Psi_{\mu \equiv\{ m_j, k_j \}, 1\leq j \leq n_s}(x,\dots,x) &= \prod_{j=1}^{n_s}\left(  \prod_{a=1}^{m_j} \frac{1+t_{j,a}}{1-t_{j,a}}\right)^x \\
 &= 
\prod_{j=1}^{n_s}\left(   \frac{\Gamma\left( -\frac{m_j}{2} + \frac{\gamma}{2} -i \frac{k_j}{\cbar}\right)\Gamma\left( \frac{m_j}{2} + \frac{\gamma}{2} +i \frac{k_j}{\cbar}\right)}
{\Gamma\left(\frac{m_j}{2} + \frac{\gamma}{2} -i \frac{k_j}{\cbar}\right)\Gamma\left( -\frac{m_j}{2} + \frac{\gamma}{2} +i \frac{k_j}{\cbar}\right)  } \         \right)^x,
\end{split}
\end{equation}
where use has been made of the identity $\Gamma( a + m )/\Gamma(a) = \prod_{k=0}^{m-1} (a+k)$. 
 Moreover, applying the same identity to the eigenvalues of the Hamiltonian (Eq. \ref{thetaMuExpr}) can also be expresssed 
 as products of Gamma functions:
\begin{equation}\label{thetaMuGamma}
\begin{split}
 \theta_{\mu \equiv\{ m_j,  k_j \}, 1\leq j \leq n_s} &=  \prod_{j=1}^{n_s} \left( \prod_{a=1}^{m_j}\frac{1}{1-t_{j,a}^2} \right)^{\frac{1}{2}}\\
 & =\prod_{j=1}^{n_s} \left( \prod_{a=1}^{m_j} \left( \frac{2}{\bar{c}}\right)^{m_j}
 \left(   \frac{\Gamma( -\frac{m_j}{2} + \frac{\gamma}{2} - i \frac{k_j}{\bar{c}} )\Gamma(  -\frac{m_j}{2} + \frac{\gamma}{2} + i \frac{k_j}{\bar{c}})}
  {\Gamma( \frac{m_j}{2} + \frac{\gamma}{2} - i \frac{k_j}{\bar{c}} )\Gamma(  \frac{m_j}{2} + \frac{\gamma}{2} +i \frac{k_j}{\bar{c}})} \right)^{\frac{1}{2}} \right).
\end{split}
\end{equation}

 The norm of such a system of string states in the thermodynamic limit was conjectured in \cite{logGammaTLD}
 to be given by:
\begin{equation}\label{normOfStringStates}
\begin{split}
 || \mu \equiv\{ m_j, k_j \}, 1\leq j \leq n_s||^2 &= \left(\sum_{j=1}^{ n_s} m_j\right)! L^{n_s} \prod_{1\leq i < j \leq n_s} \frac{4(k_i-k_j)^2 + \cbar^2(m_i+m_j)^2}{ 4 (k_i-k_j)^2 + \cbar^2(m_i - m_j)^2} \\
&\times\prod_{j=1}^{n_s}\left(\frac{m_j}{\cbar^{m_j-1}} \left(\sum_{a=1}^{m_j}\frac{1}{1-t^2_{j,a}} \right) \prod_{b=1}^{m_j } (1-t_{j,b}^2)\right).
\end{split}
\end{equation}
The formula was checked in special cases and its form is inspired by the determinantal 
 form of  the Gaudin--Korepin formula \cite{GaudinTraduction,Korepin}, to which it  reduces in the continuum limit. 
 Let us assume it holds and use of it to express the moment of  $Z_w(x,t)$ from Eq. \ref{momentOrdern}.

\subsection{Phase-space integration}
 The sum over Bethe eigenstates in Eq. \ref{momentOrdern} involves a
 sum over the Bethe states labelled by index $\mu$, which is equivalent to an integration 
over the space of rapidities. However, the string 
 states are described by non-linear functions of the rapidities, Eq. \ref{Asigma}.  The contribution 
 of a system of $n_s$ strings to the  moment of a given order $n = \sum_{j=1}^{n_s} m_j$
 is therefore expressed by  summing over each of the numbers of rapidities $m_j$, 
 and  integrating integral over each of the continuous parameters $k_j$, up to a Jacobian 
 factor worked out in Section 7.4 of \cite{logGammaTLD}. In the thermodynamic limit, this factor compensates 
 the sum of rational fractions from the norms of the string states (Eq. \ref{normOfStringStates}):
 \begin{equation}\label{phaseSpace}
 \sum_{\mu \equiv\{ m_j, k_j \}, 1\leq j \leq n_s}\longrightarrow_{L\to\infty} \prod_{j=1}^{n_s}\left( \frac{L}{2\pi} \int_{-\infty}^\infty  dk_j \sum_{a=1}^{m_j}\frac{1}{1-t_{j,a}^2}\right).
 \end{equation}
 The  net factor from the norm of  a string state with $m_j$ rapidities after this simplification 
  can therefore be expressed in terms of the string parameters in terms of  the Gamma function:
\begin{equation}\label{netFactor}
 \prod_{a=1}^{m_j}( 1 - t_{j,a}^2)^{-1}=\frac{2}{\cbar^m}\frac{\Gamma\left( -\frac{m_j}{2} + \frac{\gamma}{2} -i \frac{k_j}{\cbar}\right)
\Gamma\left( -\frac{m_j}{2} + \frac{\gamma}{2} +i \frac{k_j}{\cbar}\right)}
{\Gamma\left(\frac{m_j}{2} + \frac{\gamma}{2} -i \frac{k_j}{\cbar}\right)
\Gamma\left( \frac{m_j}{2} + \frac{\gamma}{2} +i \frac{k_j}{\cbar}\right)}.
\end{equation}

%\subsection{Analytic continuation to complex values of the order of the moments}

\section{The one-string contribution to the generating function}
 Reorganising the generating function as a sum over numbers of strings 
 in the Bethe states induces a sequence of ($\lambda$-dependent) functions of 
 as follows:
\begin{equation}\label{stringExpansion}
 g_\lambda(s) =: 1 + \sum_{n_s\geq 1} \frac{1}{n_s!}Z(n_s,s).
\end{equation}
 As in the continuum  model of directed polymers \cite{CalabreseLeDoussalFlat,CalabreseLeDoussalLong},
 the one-string term can provide a candidate expression for a kernel 
  in terms of which to write the higher-order terms in a determinantal form.
\subsection{Saddle-point approximation of the integrand}
  To express the one-string contribution to the generating function, with linear momentum $k$ (in $t_\alpha$ parametrisation),
 we need to compute the moment of order $m$, where $m$ is the number of rapidities in the string state
 as expressed \ref{momentOrdern}. The only quantity in that moment  which  has not been expressed yet is the 
 sum of the wave function over all negative values of its arguments. Let us introduce the notation
 \begin{equation}
\Omega_{w,\cbar}(m,k):=\lim_{L\to\infty}\sum_{ -L \leq y_1 \leq y_2 <\dots \leq y_m \leq 0 }e^{w\sum_{p=1}^m y_p}
\ \Psi_{\mu \equiv\{ m, k\} }(y_1,\dots, y_m)= \sum_{\sigma \in \mathcal{S}_m} A_\sigma {\mathcal{G}}^w_{\lambda_{\sigma( 1)}\dots\lambda_{\sigma( m)}},
 \end{equation}
with
\begin{equation}
{\mathcal{G}}^{w,\cbar}_{\lambda_{\sigma( 1)}\dots\lambda_{\sigma( m)}} = \lim_{L\to\infty}\sum_{ -L \leq y_1 \leq y_2 <\dots \leq y_n \leq 0 }  e^{\sum_{k=1}^m( wy_k + i\lambda_{\sigma(k)} y_k )}.
\end{equation}
 The one-string term in the generating function defined by Eq. \ref{stringExpansion} reads, with the continuation prescription of Eq. \ref{prescription}
 to complex values of the number of particles in the string (specialising Eqs \ref{PsiMuGamma},\ref{thetaMuGamma},\ref{normOfStringStates},\ref{netFactor} to the case of one string): 
\begin{equation}\label{oneStringTerm}
\begin{split}
Z(n_s = 1,s) &=   \int_C\frac{dm}{2i\pi m\sin(\pi m )}\frac{1}{\Gamma( 1 + m )} e^{-\lambda sm}\int_{-\infty}^{+\infty} dk  \;\Omega_{w,\cbar}( m,k)\\
 &\times\left( \frac{\Gamma\left( -\frac{m}{2} + \frac{\gamma}{2} -i \frac{k}{\cbar}\right)
\Gamma\left( -\frac{m}{2} + \frac{\gamma}{2} +i \frac{k}{\cbar}\right)}
{\Gamma\left(\frac{m}{2} + \frac{\gamma}{2} -i \frac{k}{\cbar}\right)
\Gamma\left( \frac{m}{2} + \frac{\gamma}{2} +i \frac{k}{\cbar}\right)  } \right)^{\frac{t}{2}+1} \times \left( \frac{\Gamma\left( -\frac{m}{2} + \frac{\gamma}{2} -i \frac{k}{\cbar}\right)\Gamma\left( \frac{m}{2} + \frac{\gamma}{2} +i \frac{k}{\cbar}\right)}
{\Gamma\left(\frac{m}{2} + \frac{\gamma}{2} -i \frac{k}{\cbar}\right)\Gamma\left( -\frac{m}{2} + \frac{\gamma}{2} +i \frac{k}{\cbar}\right)  } \right)^x.
\end{split}
\end{equation}

 Rescaling the integration variable $k$ by $\cbar$ yields
\begin{equation}\label{oneStringTerm}
Z(n_s = 1,s) =  \cbar \int_C\frac{dm}{2i\pi m\sin(\pi m )}\frac{1}{\Gamma( 1 + m )} e^{-\lambda sm}\int_{-\infty}^{+\infty} dk  \;\Omega_{w,\cbar}( m,\cbar k )\mathcal{F}(x,m,k,t+2), \\
\end{equation}
 where the time-dependent factor reads
\begin{equation}\label{timeDependentFactor}
\begin{split}
\mathcal{F}(x,m,k,t)&= \Gamma\left( -\frac{m}{2} + \frac{\gamma}{2} -i k\right)^{x + \frac{t}{2}}
\Gamma\left( -\frac{m}{2} + \frac{\gamma}{2} +i k\right)^{-x +\frac{t}{2}}\\
&\;\;\;\;\;\;\times
\Gamma\left(\frac{m}{2} + \frac{\gamma}{2} -i k \right)^{-x-\frac{t}{2} }
\Gamma\left( \frac{m}{2} + \frac{\gamma}{2} +i k\right)^{x-\frac{t}{2} }\\
&=  \left( \frac{\Gamma\left( -\frac{m}{2} + \frac{\gamma}{2} -i k\right)}{\Gamma\left(\frac{m}{2} + \frac{\gamma}{2} -i k\right)} \right)^{x +\frac{t}{2} }
  \left( \frac{\Gamma\left( -\frac{m}{2} + \frac{\gamma}{2} +i  k\right)}{\Gamma\left( \frac{m}{2} + \frac{\gamma}{2} +i k \right)} \right)^{-x +\frac{t}{2} },
\end{split}
\end{equation}
and can be studied at large time using a saddle-point approximation. Rescaling the integration  variables $m$ and $k$ by a factor 
 of $\lambda$  leads to an expression with a $\lambda$-independent exponential factor in the integrand:
\begin{equation}\label{compactOneStringTerm}
\begin{split}
Z(n_s = 1,s) = \lambda^{-1} \cbar& \int_C\frac{dm}{2i\pi m\sin(\pi m/\lambda )}\frac{1}{\Gamma( 1 + m/\lambda )} e^{-sm} \\
 &\times\int_{-\infty}^{+\infty} dk  \;\Omega_{w,\cbar}( m/\lambda,\cbar k/\lambda)\mathcal{F}(x,m/\lambda,k/\lambda,t+1), 
\end{split}
\end{equation}

%\subsection{Scaling of parameters at large time by the saddle-point method}
From the expression of the time-dependent factor in Eq. \ref{timeDependentFactor}, 
we can read off the following logarithm, whose behaviour close to $(m=0,k=0)$  is
relevant in the large-$\lambda$ limit:
\begin{equation}
 \Lambda(m,k) = \log\left(  \frac{\Gamma\left( -\frac{m}{2} + \frac{\gamma}{2} -i k\right)}{\Gamma\left(\frac{m}{2} + \frac{\gamma}{2} -i k\right)} \right).
\end{equation}
as it can be used as follows to derive a saddle-point approximation of the kernel:
\begin{equation}
\mathcal{F}(x,m,k,t)=\exp\left(  \left( x +\frac{t}{2}  \right)  \Lambda(m,k) + \left( -x +\frac{t}{2}  \right) \Lambda(m,-k)  \right). 
\end{equation}
 The logarithm $\Lambda(m,k)$ is odd in $m$, and its Taylor expansion around  the origin reads as follows, treating powers of $m$ and $k$
 as being of the same order:
\begin{equation}
 \Lambda(m,k) = \left(-\psi\left( \frac{\gamma}{2}\right)- i \psi'\left( \frac{\gamma}{2}\right) k +  \frac{1}{2}\psi''\left( \frac{\gamma}{2}\right) k^2 \right)m 
 -\frac{1}{24}\psi''\left( \frac{\gamma}{2}\right) m^3 + o(m^3),
\end{equation}
where $\psi = \Gamma'/\Gamma$ is the digamma function. 
Let us group terms that contain factors of $x$ or $t$ in the logarithm of the kernel:
\begin{equation}
\log\mathcal{F}(x,m,k,t+1)= - \psi\left( \frac{\gamma}{2}\right)mt - 2 i \psi'\left( \frac{\gamma}{2}\right) km x + \frac{1}{2}\psi''\left( \frac{\gamma}{2}\right) k^2  m t -\frac{1}{24}\psi''\left( \frac{\gamma}{2}\right) m^3 t + o(m^3). 
\end{equation}
 The first term, which is proportional both to time and to the number of particles in the string (or order of the moment before analytic continuation),
 corresponds to the additive part of the free energy. Let us discard this term from now on by shifting the energy of the configurations. This is equivalent to studying the 
  random quantity $\log{ Z_w(x,t)}+\psi(\gamma/2)t$, without changing the notations. 
 When the integration variables $m$ and $k$ in Eq. \ref{prescription} are scaled by a factor of $\lambda$ in a change of variables, the relevant logarithm 
 after shift in energy levels is therefore
\begin{equation}
\log\mathcal{F}(x,\lambda^{-1}\tilde{m},\lambda^{-1}\tilde{k},t)= - 2 i \psi'\left( \frac{\gamma}{2}\right) \tilde{k}\tilde{m}\lambda^{-2} x + \frac{1}{2}\psi''\left( \frac{\gamma}{2}\right)  \tilde{k}^2   \tilde{m} \lambda^{-3 }t -\frac{1}{24}\psi''\left( \frac{\gamma}{2}\right)  \tilde{m}^3 \lambda^{-3 }t + \dots,
\end{equation}
from which we deduce scaling laws that make the quadratic and cubic terms   independent 
 of the parameter $\lambda$.  The higher-order terms in the 
 Taylor expansion can be neglected at large time using the Laplace method, as they will all carry at least 
 a factor of $\lambda^{-5}$ from powers or $m$ and $k$, provided
\begin{equation}
  \lambda^3 = -\frac{1}{8} \psi''\left( \frac{\gamma}{2}\right)t,
\end{equation}
and
\begin{equation}
 x = \lambda^2 \frac{1}{2\psi'\left( \frac{\gamma}{2}\right)}\tilde{x}.
\end{equation}
 With these prescriptions the large-time limit coincides with the large-$\lambda$ limit, which ensures that
 the large-$\lambda$ limit of the generating function can give access to the large-time probability distribution of the 
 free energy. The Airy function $\Ai$ can be introduced  to write the  cubic term as a Laplace transform.
 Moreover  the two scale-invariant terms in the argument of the exponential function can be absorbed into the Airy function by a change of variable,
 as they are both linear in the parameter $m$.
\begin{equation}
\begin{split}
\mathcal{F}(x,\lambda^{-1}\tilde{m},\lambda^{-1}\tilde{k},t)& = \exp\left(  -i \tilde{k}\tilde{x}- 4 \tilde{k}^2\tilde{m}+ \frac{\tilde{m}^3}{3} + o(\lambda^{-1})\right) \\
& = \int_{-\infty}^\infty dy \Ai( y ) \exp\left(  ( -i \tilde{k}\tilde{x}- 4 \tilde{k}^2+ y)\tilde{m} + o(\lambda^{-1}) \right)\\
& =  \int_{-\infty}^\infty dy \Ai( y +i \tilde{k}\tilde{x}+ 4 \tilde{k}^2) \exp\left(  y \tilde{m} + o(\lambda^{-1}) \right).
\end{split}
\end{equation}
 Going back to the integral form of the one-string term in Eq. \ref{compactOneStringTerm}, we 
 need to find an equivalent of the analytic continuation of the overlap $\Omega_{w,\cbar}(m,k)$  around $(m=0,k=0)$.
Let us introduce the rescaled slope $\tilde{w}$, which will be kept fixed in the 
 large-time limit:\\
\begin{equation}
 w = \frac{\tilde{w}}{\lambda}.
\end{equation}
 At large $\lambda$ and small $m$ and $k$ the arguments of the exponential functions involved in the expression of 
 the overlap are small (the tangents of the rapidities are small, as the maximum of the real part of the $t_\alpha$ parameters is proportional to 
 $m\cbar$, Eqs \ref{Asigma},\ref{strings}):
\begin{equation}
{\mathcal{G}}^{w,\cbar}_{\lambda_1\dots\lambda_m} = \prod_{p=1}^m\frac{1}{1-e^{-(pw+  i\sum_{j=1}^p\lambda_j )}}.
\end{equation}
Even though the overlaps are calculated from integer values of $m$,
 we can trade a small value of $m$ (for which analytic continuation is needed) and a fixed value of $\cbar$, 
 for a much larger (integer) value of $m$ and a much lower value of $\cbar$, while keeping $m\cbar$ fixed.
  In the regime of small charge the rapidities are small themselves,
 and are arranged as a string in the complex plane, as Eq. \ref{Asigma} becomes:
\begin{equation}\label{strings}
t_\alpha \simeq_{\lambda_\alpha\rightarrow 0} i \lambda_\alpha/2.
\end{equation}
 The expansion at lowest non-zero order 
 in the rapidities  yields an expression analogous to that of the overlap studied
 in \cite{crossoverLD} for the crossover in the continuum  model:
\begin{equation}\label{sumForInteger}
\mathcal{G}^{\tilde{w}/\lambda,\cbar} \underset{\lambda\to\infty}{\simeq} \prod_{j=1}^m 
\frac{1}{j \frac{\tilde{w}}{\lambda}+ \frac{i}{\lambda}\sum_{a =1}^j \left( k + i\frac{\cbar}{2}\sum_{a=1}^m(m+1-2a) \right)}\left( 1 + o(1)\right),
\end{equation}
whose dominant contribution in the large-time limit coincides with the multiple integral of a plane wave in the Lieb--Liniger 
  model with a string of rapidities given by $\{k/\lambda + i\cbar/(2\lambda)(m+1 - 2a) \}_{1\leq a \leq m}$.
  In the above expression $m$ is still manifestly an integer, however the sum of the contributions of the r.h.s. weighted
 by symmetry factors 
 is known from the solution of the continuum directed polymer model to give rise to a factorised 
 form that can be analytically continued.

\subsection{Factorisation of the overlap factor in the Lieb--Liniger model}
 Let us review the factorisation of the integral over half-lines of the Bethe wave
 functions in the continuum directed polymer model \cite{CalabreseLeDoussalLong}. 
Consider a string of rapidities with $m$ elements in the  Lieb--Liniger model, 
 given in terms of a linear momentum $k$ and a charge $\cbar$ 
 by:
\begin{equation}
 \lambda_\alpha^{LL}(k,m,\cbar) = k + \frac{i\cbar}{2}( m+1 - 2\alpha),\;\;\;\;\alpha \in \{1,\dots, m\},
 \label{rapiditiesLL}
\end{equation}
 the Bethe wave function $\Psi^{LL}$ is an eigenfunction of the Lieb--Liniger Hamiltonian,
 with the following expression:
\begin{equation}
  \Psi^{LL}_{m,k,\cbar}( x_1,\dots, x_m) = \sum_{\sigma \in {\mathcal{S}}_m} A_\sigma \exp\left( i\sum_{\alpha=1}^m \lambda_{\sigma(\alpha)} x_\alpha \right),\;\;\;\;\;
A^{LL}_{\sigma}(k,m) = \prod_{1 \leq \alpha < \beta \leq m } \left(   1  + i \bar{c}\frac{\signum( x_\beta - x_\alpha + 0^+)}{\lambda_{\sigma(\alpha)} - \lambda_{\sigma(\beta)}}\right). 
\end{equation}
Moreover, the overlap integral  of this wave function with the negative half-line can be factorised as follows in the 
 case $\cbar = 1$:
\begin{equation}
 \Omega_{w,\cbar}^{LL}( k,m) = \left( \prod_{\alpha = 1}^m \int_{-\infty}^0 dy_\alpha e^{w y_\alpha} \right) \Psi_{m,k,\cbar}( y_1,\dots, y_m) = 
\sum_{P\in S_m}\left(  A^{LL}_{\sigma}(k,m)  \prod_{j=1}^m\frac{1}{jw + i\sum_{l=1}^j \lambda_{\sigma(l)}(k,m,\cbar) }   \right),
\label{sumForm}
\end{equation}
can be factorised as follows in the  case $\cbar = 1$ (see Section 5 in \cite{crossoverLD}):
\begin{equation}
 \Omega_{w,\cbar=1}^{LL}( k,m,\cbar=1, w )  = \frac{m!}{i^n \prod_{\alpha=1}^m(\lambda_\alpha(k,m,\cbar = 1 )  -iw)}\prod_{1\leq \alpha<\beta \leq m}
\frac{i+\lambda_\alpha(k,m,\cbar = 1 )  + \lambda_\beta(k,m,\cbar = 1 )  -2iw}{\lambda_\alpha(k,m,\cbar = 1 )  + \lambda_\beta(k,m,\cbar = 1 )  -2iw},
\end{equation}
which upon substituting the string of rapidities described in Eq. \ref{rapiditiesLL} yields
\begin{equation}
\Omega_{w,\cbar=1}^{LL}( k,m )  = \frac{(-1)^m \Gamma( 2ik + 2w )}{\Gamma( 2ik + 2w+m)}.
\end{equation}

To restore the value of $\cbar$, let us start from Eq. \ref{sumForm} by factorising one power of $\cbar$ per factor in each of the 
 terms in the sum:
\begin{equation}
\begin{split}
 \Omega_{w,\cbar}^{LL}( k,m ) &= \cbar^{-m} \sum_{P\in S_m}\left(  A^{LL}_{P}(k,m)  \prod_{j=1}^m\frac{1}{j\frac{w}{\cbar}+ i\sum_{l=1}^j \lambda_{P(l)}\left(\frac{k}{\cbar},m,\cbar = 1\right) }   \right)\\
& = \frac{m!}{(i\cbar)^m \prod_{\alpha=1}^m\left(\lambda_\alpha(\frac{k}{\cbar},m,\cbar = 1 )  -i\frac{w}{\cbar}\right)}\prod_{1\leq \alpha<\beta \leq m}
\frac{i+\lambda_\alpha(\frac{k}{\cbar},m,\cbar = 1 )  + \lambda_\beta(\frac{k}{\cbar},m,\cbar = 1 )  -2i\frac{w}{\cbar}}{\lambda_\alpha(\frac{k}{\cbar},m,\cbar = 1 )  + \lambda_\beta(\frac{k}{\cbar},m,\cbar = 1 )  -2i\frac{w}{\cbar}}.
\end{split}
\end{equation}
% The overlap of the Lieb--Liniger is therefore expressed as: 
%\begin{equation}
%\Omega^{LL}( k,m,\cbar, w )  = \frac{(-1)^m \Gamma\left( \frac{2ik + 2w}{\cbar} \right)}{\cbar^m\Gamma\left( \frac{2ik + 2w}{\cbar}+m\right)}.
%\end{equation}

\subsection{Large-time limit of the one-string contribution}
 The equivalent of the overlap factor in the large-time limit identified in \ref{sumForInteger} from the lowest non-zero 
 order expansion in the rapidities for integer $m$ can therefore be analytically continued 
  by substituting  the above Lieb--Liniger expression: 
\begin{equation}
\Omega_{\tilde{w}/\lambda,\cbar}( m,k/\lambda ) \underset{m\rightarrow 0}{\simeq} \frac{(-1)^m \Gamma\left( \frac{2ik + 2\tilde{w}}{\lambda\cbar} \right)}{\cbar^m\Gamma\left( \frac{2ik + 2\tilde{w}}{\lambda\cbar}+m\right)}.
\end{equation}
 It should be noted that the occurrence of this expression from the Lieb--Liniger model comes 
 merely from the large-time limit and the continuation to complex values of $m$, while $\gamma$ is kept 
 fixed.  The lattice spacing is still equal to one. 
 In the continuum limit will be taken eventually, $\gamma$ will go to infinity as the lattice spacing will go to zero.\\

 For a fixed value of the rescaled slope $\tilde{w}$ and a large value of time,
 the asymptotic form of the analytic continuation of the integral over the negative real axis can be substituted into
 the expression of the  one-string contribution expressed in Eq. \ref{compactOneStringTerm}, together 
 with the integral representation $m^{-1} = \int_0^\infty e^{-mv}$,   to yield:
\begin{equation}
\begin{split}
Z(n_s = 1,s)& \sim_{\lambda\infty}\int_0^\infty dv \int_{-\infty}^\infty dk \int_{-\infty}^\infty dy  \int_C\frac{dm}{2i\pi \sin(\pi m/\lambda )}\frac{1}{\Gamma( 1 + m/\lambda )}\\
 &\int_{-\infty}^{+\infty} dk  \;  \frac{\Gamma\left(\frac{2(ik+\tilde{w})/\cbar}{\lambda}\right)}{\Gamma\left(\frac{(2(ik+\tilde{w})/\cbar+m)}{\lambda}\right)}  
\times \Ai(y +i k x+ 4 k^2) \exp\left(   -mv -ms + my \right).\\
& \sim_{\lambda\infty} \int_0^\infty dv \int_{-\infty}^\infty dk\int_{-\infty}^\infty dy  \int_C\frac{dm}{2i\pi \sin(\pi m/\lambda )}\\
 &\int_{-\infty}^{+\infty} dk  \;  \frac{\Gamma\left(\frac{2(ik+\tilde{w})}{\lambda}\right)}{\Gamma\left(\frac{2(ik+\tilde{w}+m)}{\lambda}\right)}  \times \Ai(y+ v + s +i k \tilde{x}+ 4 k^2) \exp\left(  my \right),
\end{split}
\end{equation}
 in which a shift of the integration variable $y$ has been used to transfer the variables $s$ to the Airy function.
 We can simplify the integrand in the large-time limit by using the equivalent  $\Gamma( u/\lambda )\sim_{\lambda\to\infty}\lambda / u$
\begin{equation}\label{limitOneString}
\begin{split}
Z(n_s = 1,s)& \sim_{\lambda\infty}\int_0^\infty dv \int_{-\infty}^\infty dk\int_{-\infty}^\infty dy  \int_C\frac{dm}{2i\pi m} \frac{(2ik + 2\tilde{w})/\cbar + m}{ (2ik + 2\tilde{w})/\cbar }  \times \Ai(y+ v + s +i k \tilde{x}+ 4 k^2) e^{  my}\\
 & \sim_{\lambda\infty}\int_0^\infty dv \int_{-\infty}^\infty dk\int_{-\infty}^\infty dy  \left(  -2\theta(y) - \frac{\cbar}{ ik + \tilde{w}}\delta(y)\right)  \times \Ai(y+ v + s +i k \tilde{x}+ 4 k^2),
\end{split}
\end{equation}
where the Laplace transforms of the Dirac mass and step function $\theta(y) = \mathbf{1}( y> 0)$ 
 have been used to integrate over $m$.

\section{Large-time limit of the generating function}

\subsection{Determinantal form of the higher-order terms in the string expansion}
 As in the Lieb--Liniger case, assuming the generating function has the structure 
 of a Fredholm determinant, we can read off the kernel from the one-string calculation.
  Introducing the kernel
\begin{equation}
  K_{\gamma,s}(v_1,v_2) = \int_{-\infty}^\infty dk\int_{-\infty}^\infty dy  \left(  2\theta(y) + \frac{4}{(\gamma-1)( ik + \tilde{w})}\delta(y)\right)  \times \Ai(y+ v_1+v_2 + s +i k \tilde{x}+ 4 k^2) e^{-2ik( v_1-v_2)},
\end{equation}
  the large-time limit of the one-string term reduces to its trace:
 \begin{equation}
 Z(n_s = 1,s) \sim_{\lambda\infty} -\int_{-\infty}^\infty  K_s(v,v) dv.
\end{equation}
 In order to rewrite the generating function in terms of the above kernel, we have to 
 insert  factors from the multiple-string contributions to the norm
 of the string states.\\

 Going back to the formal expression of the moment of order $n$, 
 the inter-string factors from the norms of the string states (Eq. \ref{normOfStringStates}) yield the 
  starting formula for the $n_s$-string contribution (while thanks to the expression of the eigenvalues, the time-evolution 
  factor reduces to the product of the contribution of each string): 
\begin{equation}
\begin{split}
 Z^{mom}(n_s,s) =& \sum_{m_1,\dots,m_{n_s} = 1}^\infty\prod_{j=1}^{n_s}\left( \frac{2^{m_j}}{m_j}\int_{-\infty}^\infty\frac{dk_j}{2\pi} \mathcal{F}(x,m_j,k_j,t+2) \right)  \\
 &\times \Omega_{w,\cbar}(\{ m_j, \cbar k_j \}_{1\leq j \leq n_s})
 \prod_{1\leq i<j\leq n_s}\frac{4(k_i-k_j)^2 +(m_i-m_j)^2}{4(k_i-k_j)^2 + (m_i+m_j)^2}).
\end{split}
\end{equation}  
The inter-string factors from the norm allow to use the same 
 crucial identity as in \cite{CalabreseLeDoussalLong,logGammaTLD} 
\begin{equation}
\prod_{1\leq i<j\leq n_s}\frac{4(k_i-k_j)^2 +(m_i-m_j)^2}{4(k_i-k_j)^2 + (m_i+m_j)^2} = \det\left(\left( \frac{1}{2i(k_i-k_j) + m_i + m_j} \right)\right)|_{n_s\times n_s}\prod_{j=1}^{n_s}(2m_j),
\end{equation}
 to give the $n_s$-string contribution the form of a determinant. Applying the integration prescription of Eq. \ref{prescription} $n_s$ times 
 leads to an expression in which each of the real  integration variables $k_1,\dots,k_{n_s}$ has been rescaled:
\begin{equation}
\begin{split}
Z( n_s, s) =& \left( \prod_{j=1}^{n_s} \int_{C_j}\frac{-dm_j}{2i\pi \sin(\pi m_j) } 2^{m_j+1}  \int_{-\infty}^\infty\frac{dk_j}{2\pi} \mathcal{F}(x,m_j,k_j/\lambda,t+2) e^{-\lambda m_j s}\right)\\ 
&\times \Omega_{w,\cbar}(\{ m_j, \cbar k_j /\lambda\}_{1\leq j \leq n_s})  \det\left(\left( \frac{1}{2i(k_i-k_j) + \lambda(m_i + m_j)} \right)\right)|_{n_s\times n_s}
\end{split}
\end{equation}  
 As the maximum rapidity in a system of strings is bounded by the maximum of $m_j\cbar$, we 
  use a small-rapidity expansion of the overlap integrals in each of the strings, which allows us to 
make us of the factorisation of the overlap of the wavefunction in the Lieb--Liniger model:\\
\begin{equation}
\begin{split}
\Omega_{w,\cbar}(\{ m_j, \cbar k_j /\lambda\}_{1\leq j \leq n_s} )&\simeq_{\lambda\to\infty} \left( \cbar^{-m_j }\prod_{j=1}^{n_s} \frac{(-1)^{m_j}\Gamma(z_{jj}/\lambda)}{\Gamma((z_{jj} + m_j)/\lambda)}\right)\\
&\times  \prod_{1\leq i<j\leq n_s} \frac{\Gamma( 1 -  (z_{ij} + (m_i+m_j)/2)/\lambda)\Gamma( 1 -  (z_{ij} - (m_i+m_j)/2)\lambda )}
{\Gamma( 1 -  (z_{ij} + (m_j-m_i)/2)/\lambda)\Gamma( 1 -  (z_{ij} - (m_j-m_i)/2)/\lambda)}
\end{split}
\end{equation}
where
\begin{equation}
 z_{ij}=  i( k_j+k_j) + 2 w, \;\;\;i,j \in \{ 1,\dots, n_s\}.
\end{equation}
 The inter-string factors coming from the overlap integral  go to zero in the large-time 
 limit once the variables are rescaled. Inserting the large-$\lambda$ expansion of the time-dependent factor with the scaling 
 of $\lambda$ identified at the one-string level, together with the representation 
  of the determinant  as a sum over permutations, and an integral representation of 
 each of the factors, reproduces all the algebraic steps of the derivation of the 
 Fredholm determinant in the continuum model: 
\begin{equation}
Z (n_s,s) \simeq_{t\to \infty}(-1)^{n_s} \left(  \left( \prod_{j=1}^{n_s}  \int_{0}^\infty dv_j\right) \det K_{\gamma,s}(v_i,v_j)|_{n_s\times n_s}\right)(1 + o(1) ) 
\end{equation}
 
 Having derived the $n_s$ by $n_s$  determinantal form of the $n_s$-string 
 in the large-time limit, we can use the identities obtained in Section 9 of \cite{crossoverLD} through variations on formulas
 involving integrals of products of Airy functions \cite{AiryBook}, up to a factor of $\cbar = 4/(\gamma-1)$ in front 
 of the $\delta$-function in the kernel, and obtain the Fredholm determinant:
\begin{equation}
 g_\infty( s )= \mathrm{Det}\left( 1 - \mathcal{K}_{\gamma,s}\right)
\end{equation}

\begin{equation}
\begin{split}
 \mathcal{K}_{\gamma,s}&(v_1,v_2)= \theta(v_1) \theta(v_2)\left(  \int_0^\infty dy \Ai\left( y+v_1+ 2^{-2/3}\left(  s + \frac{\tilde{x}^2}{16} \right)\right)\Ai\left( y+ v_2 +  2^{-2/3}\left(  s + \frac{\tilde{x}^2}{16}  \right)\right) \right)\\
 & +\frac{4\theta(v_1) \theta(v_2)}{\gamma-1}\left(-\int_{-\infty}^0 dy \Ai( v_1 +\sigma+ y) \Ai( v_2+\sigma-y) e^{2y u} + 2^{-1/3}\,\Ai\left( v_1 + v_2 + 2\sigma - 2u^2\right) e^{-u(v_1-v_2)}\right),
\end{split}
\end{equation}
 with 
\begin{equation}
 u =- 2^{-2/3}\left(  \tilde{w} + \frac{\tilde{x}^2}{16}\right), \;\;\;{\mathrm{and}} \;\;\;\sigma = 2^{-2/3}\left(  s + \frac{\tilde{x}^2}{16} \right).
\end{equation}
\subsection{GUE  limit}
 The second part of the kernel vanishes in the  limit 
\begin{equation}
\tilde{w} + \frac{\tilde{x}}{8} \to \infty.
\end{equation}
 If we let time go to infinity while keeping $s+\tilde{x}^2/16$ finite, the coordinate $x$ scales 
 with time according to the saddle-point approximation so that it stays small 
 in scale of time
 \begin{equation}
 \frac{x}{t} \simeq_{t\to\infty} O(t^{-1/3}).
 \end{equation}
  We are therefore close to the diagonal of the square lattice (in the $(i,j)$ coordinates described in the intrduction, because $x$ 
 is proportional to the difference $i-j$), which is the sector 
  of the model with boths fixed ends for which explicit expressions for the elastic constant were worked out in 
 \cite{logGammaTLD} in terms of the model parameter $\gamma$.
 Moreover, we can read off the relevant expression in the argument 
 of the Airy kernel in terms of the coordinates $x$ and $t$ as
 \begin{equation}
 \lambda \frac{\tilde{x}^2}{16} = 2\frac{\psi'\left( \frac{\gamma}{2}\right)^2}{\psi''\left( \frac{\gamma}{2}\right)}\frac{x^2}{t},
\end{equation}
so that if the random variable $\chi_2$ has the GUE distribution $F_2$ as its cumulative distribution 
 function, the following identity holds in law
\begin{equation}\label{parabolicShift}
 -\lambda f = \lambda \chi_2 +\kappa\frac{x^2}{4t},
\end{equation}
with 
\begin{equation}
 \kappa = -8\frac{\psi'\left( \frac{\gamma}{2}\right)^2}{\psi''\left( \frac{\gamma}{2}\right)}.
\end{equation}
 The parabolic term is Eq. \ref{parabolicShift} identical to the one worked out in the
 diagonal region in the model with both ends fixed (Section 10 in \cite{logGammaTLD}).
 It is known  to reduce to the droplet solution in the continuum limit, which corresponds to 
  the GUE behaviour of the large-time limit of the model with both ends fixed in the diagonal sector.
% Moreover this term is invariant under divisin of $x$ and $t$ by a factor of $\cbar$ and $\cbar^2$ 
 % respectively.

\subsection{GOE limit}

To capture the GOE limit we need to send $\tilde{w} + \tilde{x}/8$ to $-\infty$ 
 while keeping $\tilde{w}$ positive, in which limit only the last term in the kernel contributes:
\begin{equation}
 \mathcal{K}_s(v_1,v_2)  \simeq_{\tilde{w} + \tilde{x}/8 \to -\infty}\theta(v_1) \theta(v_2) \Ai\left( v_1 + v_2 + 2\sigma - 2u^2\right) ,
\end{equation}
 Scaling the coordinates $x$ and and $t$ by a factor of   $\cbar$ and $\cbar^2$  respectively 
 restores the scale $\cbar$ in the statements  of \cite{crossoverLD},
 and denoting by $\chi_1$ the random variable that has the GUE distribution $F_2$ as its cumulative distribution 
 function we read off the identity in law:
 \begin{equation}
 -\lambda f = \lambda \chi_1 + \lambda\left( 4\frac{\tilde{w}^2}{\cbar^2} +  \tilde{w}\frac{\tilde{x}}{\cbar} \right).
\end{equation}
Working out the last two terms in terms of the coordinates $x$, $t$ and the parameter
 of the model:
 \begin{equation}
  4\frac{\lambda^3}{\cbar^2}\tilde{w}^2+\lambda  \tilde{w}\frac{\tilde{x}}{\cbar} = - \frac{1}{2} \psi''\left( \frac{\gamma}{2}\right)w^2 \frac{t}{c^2}+ 2\psi'\left( \frac{\gamma}{2}\right)w\frac{x}{\cbar},
\end{equation}
we can take the continuum limit by introducing a lattice spacing $\aStep$ sent to zero with fixed Lieb--Liniger parameter $\cbar_{LL}$  through the prescription
\begin{equation}
\cbar = \aStep \cbar_{LL},\;\;\;   \gamma = 1 + \frac{4}{\aStep \cbar_{LL}}\to\infty
\end{equation}
where the factor of $1/8$ ensures that the factors $\theta_\mu^t$ reduce to the
 time-evolution factor of the Lieb--Liniger model in the continuum limit (see Section 5 of \cite{logGammaTLD}).
 The asymptotic expansion  of the digamma function
\begin{equation}
  \psi(u)\simeq_{u\to\infty}= \log u -\frac{1}{2u}- \frac{1}{12 u^2} + O(u^{-4})
\end{equation}
 yields
\begin{equation}
 4\frac{\lambda^3}{\cbar^2}\tilde{w}^2+\lambda  \tilde{w}\frac{\tilde{x}}{\cbar} \simeq_{LL} -\frac{t}{8}w^2 +  w x = w_{LL}x_{LL}+t_{LL}w_{LL}^2,
\end{equation}
 where  the continuum  coordinates $x_{LL}$ and $t_{LL}$ are scaled by powers of the lattice spacing
\begin{equation}
x = \frac{x_{LL}}{\aStep},\;\;\;\; t = 8\frac{t}{\aStep^2},\;\;\;\;w = \aStep w_{LL},
\end{equation}
where the factor of $8$ ensures that the factors $\theta_\mu^t$ reduce to the
 time-evolution factor of the Lieb--Liniger model in the continuum limit (see Section 5 of \cite{logGammaTLD}),
 and the scaling of the slope $w_{LL}$ is induced by prescribing that the shift of the free energy $wx$ should 
 be fixed in the continuum limit. The shift in the free energy therefore coincides with the one in the GOE limit 
 of the crossover in the KPZ equation.

\end{document}